\documentclass[11pt]{article}
\usepackage{amsmath,amssymb}  % Better maths support & more symbols
\usepackage{graphicx}
\usepackage{bm}  % Define \bm{} to use bold math fonts
\usepackage{authblk}
\numberwithin{equation}{section}
\begin{document}
\title{ Reversible diffusion-influenced reactions of an isolated pair on two dimensional surfaces}
\author[1]{Thorsten Pr\"ustel} 
\author[2]{M. Tachiya} 
\affil[1]{Laboratory of Systems Biology\\National Institute of Allergy and Infectious Diseases\\National Institutes of Health}
\affil[2]{National Institute of Advanced Industrial Science and Technology}
\maketitle
\let\oldthefootnote\thefootnote 
\renewcommand{\thefootnote}{\fnsymbol{footnote}} 
\footnotetext[1]{Email: prustelt@niaid.nih.gov, m.tachiya@aist.go.jp} 
\let\thefootnote\oldthefootnote 
\abstract
{We investigate reversible diffusion-influenced reactions of an isolated pair in two dimensions.
%The Green's function for this case has been obtained before by 
To this end, we employ convolution relations that permit deriving the survival probability of the reversible reaction directly in terms of the survival probability of the
irreversible reaction. Furthermore, we make use of the mean reaction time approximation to write the irreversible survival probability in restricted spaces as a single exponential.
In this way, we obtain exact and approximative expressions in the time domain for the reversible survival probability for three different two dimensional diffusion spaces: 
The infinite and restricted plane as well as the surface of a sphere.
Our obtained results should prove useful in the context of membrane-bound reversible diffusion-influenced reactions in cell biology.
}
\section{Introduction}
Diffusion-influenced reactions \cite{Rice:1985} can be modeled by solutions of the Smoluchowski equation in the presence of certain types of boundary conditions (BC).
Among all solutions, Green's functions (GF) are of particular importance, because they permit constructing the solution for any given initial distribution
and they can be used to calculate other relevant quantities, for instance, the survival probability \cite{Agmon:1984, kimShin:1999, TPMMS_2012JCP}. However, in most cases, the GF is not directly observable, but its mathematical derivation can be somewhat involved. Therefore, it is worth while looking for alternative approaches that dispense with the GF of the Smoluchowski equation altogether
and still permits the calculation of observable quantities in a less labor intense way. 

One example of such an approach is provided by convolution relations \cite{Tachiya:1980, Agmon:1990p10} that relate the survival probability of the reversible reaction to the survival probability of the
irreversible reaction. The advantage of this method lies in the fact that the survival probability of the irreversible reaction is already known for a number of cases. Hence, the reversible survival probability can be calculated directly in terms of the irreversible survival probability, without the need to solve the corresponding complicated initial-boundary problem of the Smoluchowski equation. 
Even if the irreversible survival probability is not known, it can be derived as the solution of its equation of motion, the Sano-Tachiya (S-T) equation \cite{Sano_Tachiya:1979}, which is also more easily solved than the underlying Smoluchowski equation.

The convolution relations provide another advantage. In the case of restricted diffusion spaces, the survival probability of the irreversible reaction is typically given by a sum of exponentials with various decay times. By means of the
so-called mean reaction time approximation \cite{Sano_Tachiya:1981}, and except at very short times, the survival probability can be written as a single exponential with a characteristic decay time that can be identified with the mean reaction time. Combining these approximative expressions for the irreversible survival probability with the convolution relations immediately generates approximative expressions for the reversible survival probability as well.

In this paper, we will use these methods to study the reversible diffusion-influenced reaction of an isolated pair in two dimensions (2D). 
Diffusion in 2D is of particular interest for several reasons.
Conceptually, it is distinct because 2D is the critical dimension regarding recurrence and transience of random walks \cite{Toussaint_Wilczek_1983} and because the steady-state solution of the diffusion equation is inconsistent with the boundary condition at infinity \cite{Emeis_Fehder_1970}.
In view of biological applications, the theoretical treatment of the 2D case facilitates a better understanding of reversible membrane-bound reactions in cell biology, shedding new light on phenomena like
signal-induced heterogeneities and receptor clustering \cite{bethani:2010}.
Finally, from a more technical point of view, the mathematical treatment of the 2D case appears more cumbersome in general than its 1D and 3D counterparts \cite{TPMMS_2012JCP, TPMMS_2013JCP}, emphasizing the need for less labor intense methods. 

The paper is organized as follows. After introducing the general theoretical context, we will discuss the central convolution relations. We will use these relations to
obtain exact and approximative solutions for three different 2D diffusion spaces: 
The infinite and restricted plane as well as the surface of a sphere. Although the case of the infinitely extended plane was already solved by using a GF of the 2D Smoluchowski equation \cite{TPMMS_2012JCP}, the cases of the restricted plane and the spherical surface have not yet been solved. Here we solve all the three cases by using the convolution relations. Finally, we will compare and discuss the results.
In the appendix, we will present an alternative derivation of the convolution relations.

\section{Theory}
\subsection{Smoluchowski equation}
We consider an isolated pair of two disklike molecules $A$ and $B$ with diffusion constants $D_{A}$ and $D_{B}$, respectively.
The molecules may bind when their separation equals the encounter distance $a$ to form a bound molecule $AB$. In the case of a reversible reaction, the bound molecule may dissociate to form an unbound pair $A+B$ again.
This system can equivalently be described as the diffusion of a point-like particle with diffusion constant $D=D_{A}+D_{B}$ around a static disk with radius $a$. Reactions are introduced by imposing boundary conditions at the disk's surface.  
The probability density function (PDF) $p(r, t\vert r_{0})$ yields the probability to find the particles at a separation $r$ at time $t$, provided that the separation was initially $r_{0}$ at time $t=0$. We assume that the system is centrosymmetric and that the interaction potential vanishes. Then, the time evolution of $p(r, t\vert r_{0})$ is governed by the Smoluchowski equation \cite{smoluchowski:1917}   
\begin{equation}\label{diffusionEq}
\frac{\partial}{\partial t}p(r, t\vert r_{0}) = D\bigg(\frac{\partial^{2}}{\partial r^{2}} + \frac{1}{r}\frac{\partial}{\partial r}\bigg)p(r, t\vert r_{0}), \quad r\geq a.
\end{equation}
The initial condition is
\begin{equation}\label{initial_bc}
2\pi r_{0} p(r, 0\vert r_{0})=\delta(r-r_{0}).
\end{equation}
In case of the infinitely extended plane, one requires the boundary condition
\begin{equation}\label{inf_bc}
\lim_{r \rightarrow\infty} p(r, t\vert r_{0}) = 0,
\end{equation}
while for the restricted plane, one imposes the BC
\begin{equation}\label{res_bc}
\frac{\partial}{\partial r}p(r, t\vert r_{0})\big\vert_{r=b} = 0, \quad b > a.
\end{equation}
Regardless of the underlying diffusion space, the GF describing the diffusion-influenced reaction is only defined for $r \geq a > 0$ and one has to specify a BC for $r=a$ that takes into account the behavior at the encounter distance.
The irreversible association reaction is taken into account by the radiation BC \cite{collins1949diffusion} that is characterized by an intrinsic association constant $\kappa_{a}$.
\begin{eqnarray}\label{RadBC}
-J(a,t\vert r_{0}) &=& \kappa_{a}p(a, t\vert r_{0}).
\end{eqnarray}
Here, $J(r,t\vert r_{0})$ refers to the total diffusional flux
\begin{eqnarray}\label{defFlux}
-J(r,t\vert r_{0}) = 2\pi r D \frac{\partial}{\partial r}p(r, t\vert r_{0}).
\end{eqnarray}
To describe reversible reactions, the radiation BC has to be generalized to the back-reaction BC that involves additionally an intrinsic dissociation constant $\kappa_{d}$ \cite{Agmon:1984, kimShin:1999, TPMMS_2012JCP}. 
\begin{eqnarray}\label{BRBC}
-J(a,t\vert r_{0}) &=& \kappa_{a}p(a, t\vert r_{0}) - \kappa_{d}[1-S(t\vert r_{0})].
\end{eqnarray}
Note that for $\kappa_{d}=0$, Eq.~\eqref{BRBC} reduces to the radiation BC Eq.~\eqref{RadBC}, as it should.

Knowledge of the GF permits to derive further important quantities, in particular the survival probability $S(t\vert r_{0})$, i.e. the probability that a pair of molecules with initial distance $r_{0}$ 
is separated at time $t$
\begin{eqnarray}\label{defS}
S(t\vert r_{0})  =  2\pi\int^{b}_{a} p(r, t\vert r_{0}) r dr, 
\end{eqnarray}
where $b = \infty$ in the case of the infinitely extended plane.

So far, the theory was formulated for the initially unbound state, but quite analogously one can treat the case of an initially bound state. The corresponding PDF is denoted by $p(r, t\vert\ast)$ and 
it also satisfies Smoluchowski equation Eq.~(\ref{diffusionEq}). However, the initial condition is now
\begin{equation}\label{initialBCBound}
p(r, t=0\vert\ast) = 0,
\end{equation}  
and the back-reaction BC becomes
\begin{eqnarray}\label{BRBCBound}
-J(a,t\vert \ast)  = \kappa_{a}p(a, t\vert \ast) - \kappa_{d}[1-S(t\vert \ast)].
\end{eqnarray} 
The "survival probability" $S(t\vert \ast)$ is the probability that an initially bound pair of molecules is found unbound at time $t$
\begin{eqnarray}\label{defSbound}
S(t\vert \ast)  & := & 2\pi\int^{\infty}_{a} p(r, t\vert \ast) r dr \label{defSBound}
\end{eqnarray}
The notion of a "survival probability" is somewhat misleading in the reversible case. For the irreversible reaction, a non-vanishing probability to observe an unbound isolated pair at a time $t$ necessarily means that no association reaction has occurred before. For the reversible reaction, by contrast, association and subsequent dissociation events may occur possibly many times before $t$ and hence contribute to the probability of finding an isolated pair separated at a time $t$. Therefore, the quantities $S(t\vert r_{0})$, $S(t\vert \ast)$ could be more precisely referred to as a separation probability of the initially unbound and bound pair, respectively \cite{Agmon:1990p10}. Nevertheless, the term "survival probability" has been commonly used in the literature \cite{kimShin:1999} also for the reversible case. In the following, we will use survival probability interchangeably with separation probability. 

We would like to point out that we will make use of the notation $p_{\text{irr}}(r,t\vert r_{0}), S_{\text{irr}}(t\vert r_{0})$ and $p_{\text{rev}}(r,t\vert r_{0}), p_{\text{rev}}(r,t\vert \ast), S_{\text{rev}}(t\vert r_{0})$, $S_{\text{rev}}(t\vert \ast)$ to refer
to the GF and survival probability in the case of the irreversible reaction and to those in the case of the reversible reaction, respectively. 
The GF and survival probability in the case of the irreversible reaction are assumed to satisfy a radiation BC Eq.~\eqref{RadBC}.
\subsection{Survival probabilities and convolution relations}
Instead of calculating the GF corresponding to the quite complicated initial and boundary value problem specified by Eqs.~\eqref{initial_bc},~\eqref{BRBC} and Eqs.~\eqref{initialBCBound},~\eqref{BRBCBound} for the initially unbound and bound state, respectively,
we will employ convolution equations relating $S_{\text{rev}}(t\vert \ast)$ to $S_{\text{irr}}(t\vert r_{0})$
\begin{eqnarray}\label{eq1}
S_{\text{rev}}(t\vert \ast) &=& \kappa_{d}\int^{t}_{0}P_{\text{rev}}(\tau\vert \ast)S_{\text{irr}}(t-\tau\vert a) d\tau,\\
\frac{\partial P_{\text{rev}}(t\vert \ast)}{\partial t} &=&  -\kappa_{d}P_{\text{rev}}(t\vert \ast) - \kappa_{d}\int^{t}_{0}P_{\text{rev}}(\tau\vert \ast)\frac{\partial S_{\text{irr}}(t-\tau\vert a)}{\partial t} d\tau,\label{eq2}\qquad\qquad
\end{eqnarray}
where $P_{\text{rev}}(t\vert \ast)$ denotes the fraction of the bound state at time $t$
\begin{equation}
P_{\text{rev}}(t\vert \ast) = 1 - S_{\text{rev}}(t\vert \ast).
\end{equation}
 Eqs.~\eqref{eq1},~\eqref{eq2} have been already given in Ref.~\cite{Tachiya:1980}, see also \cite{Agmon:1990p10}. Note that in Ref.~\cite{Tachiya:1980} the convolution equations are actually formulated in a more general form that also takes into account the possibility of a decay of the bound state according to the reaction scheme $A+B\leftrightharpoons C \rightarrow D$.

The convolution relations Eqs.~\eqref{eq1},~\eqref{eq2} can be derived as follows. 
First, we consider Eq.~\eqref{eq1}.
The amount of the bound state which is converted to the unbound state between time $\tau$ and $\tau + d\tau$ is given by $\kappa_{d}P_{\text{rev}}(\tau\vert\ast)d\tau$. When the bound state is converted to the unbound state, the latter is assumed to be formed with the two constituent reactants separated by the reaction distance $a$. The probability that the unbound state which is formed at time $\tau$ with the two constituent reactants separated by $a$ is still alive at time $t$ is given by $S_{\text{irr}}(t-\tau\vert a)$. Therefore, the fraction of the unbound state is given by Eq.~\eqref{eq1}.

Let us now turn to Eq.~\eqref{eq2}.
The first term on the rhs represents the decay of the bound state between $t$ and $t+dt$. The second term represents the formation of the bound state between $t$ and $t+dt$. It is calculated in the following way. The amount of the bound state which is converted to the unbound state between time $\tau$ and $\tau + d\tau$ is given by $\kappa_{d}P_{\text{rev}}(\tau\vert\ast)d\tau$. When the bound state is converted to the unbound state, the latter is formed with the two constituent reactants separated by the reaction distance $a$. The probability that the unbound state which is formed at time $\tau$ with the two constituent reactants separated by $a$ is still alive at time $t$ is given by $S_{\text{irr}}(t-\tau\vert a)$. When the unbound state decays, the bound state is formed. The rate at which the unbound state formed between time $\tau$ and $\tau + d\tau$ with the two constituent reactants separated by $a$ will decay to form the bound state at time $t$ is given by $-\partial S_{\text{irr}}(t-\tau\vert a)/\partial t$. Therefore, the formation rate of the bound state is given by the second term on rhs of Eq.~\eqref{eq2}.

There are analogue convolution equations for the initially unbound state
\begin{eqnarray}\label{eq1unbound}
S_{\text{rev}}(t\vert r_{0}) &=& S_{\text{irr}}(t\vert r_{0}) + \kappa_{d}\int^{t}_{0}P_{\text{rev}}(\tau\vert r_{0})S_{\text{irr}}(t-\tau\vert a) d\tau,\\
\frac{\partial P_{\text{rev}}(t\vert r_{0})}{\partial t} &=& 
 -\kappa_{d}P_{\text{rev}}(t\vert r_{0}) \nonumber\\  
&&- \kappa_{d}\int^{t}_{0}P_{\text{rev}}(\tau\vert r_{0})\frac{\partial S_{\text{irr}}(t-\tau\vert a)}{\partial t} d\tau - \frac{\partial S_{\text{irr}}(t\vert r_{0})}{\partial t},\label{eq2unbound}\qquad\qquad
\end{eqnarray}
where 
\begin{equation}
P_{\text{rev}}(t\vert r_{0}) = 1 - S_{\text{rev}}(t\vert r_{0}).
\end{equation}
Eqs.~\eqref{eq1unbound} and \eqref{eq2unbound} can be derived by a quite analogous line of reasoning as presented above for Eqs.~\eqref{eq1} and ~\eqref{eq2}.
In addition, in the appendix we give an alternative derivation of the convolution relations. 

One of the virtues of the convolution relations Eqs.~\eqref{eq1} and ~\eqref{eq1unbound} is that they can be easily solved in the Laplace domain 
\begin{eqnarray}
\tilde{S}_{\text{rev}}(s\vert\ast) &=& \frac{\kappa_{d}}{s}\frac{\tilde{S}_{\text{irr}}(s\vert a)}{1+\kappa_{d}\tilde{S}_{\text{irr}}(s\vert a)}, \label{eq1Laplace}\\
\tilde{S}_{\text{rev}}(s\vert r_{0}) &=& \frac{1}{s} + \frac{\tilde{S}_{\text{irr}}(s\vert r_{0}) -s^{-1} }{1+\kappa_{d}\tilde{S}_{\text{irr}}(s\vert a)}.\label{eq2Laplace}
\end{eqnarray}
The survival probability $S_{\text{irr}}(t\vert r_{0})$ describing the irreversible case is already known for a number of different cases. Otherwise, $S_{\text{irr}}(t\vert r_{0})$can be found as the solution of the S-T equation.
In the following, we will exploit this fact to calculate
$S_{\text{rev}}(t\vert\ast), S_{\text{rev}}(t\vert r_{0}) $ in a less labor intense way than via the use of GF. 
\section{Calculation of the survival probability}
\subsection{Infinitely extended plane}
First, we consider an isolated pair reversibly reacting in the infinitely extended plane. The survival probability in the irreversible case is known, see Ch.~13.5., Eq.~(13) in Ref.~\cite{carslaw1986conduction}
\begin{equation}\label{SradLaplace}
s\tilde{S}_{\text{irr}}(s\vert r_{0}) = 1 - \frac{hK_{0}(qr_{0})}{qK_{1}(qa) + hK_{0}(qa)},
\end{equation}
where $K_{0}, K_{1}$ refer to the modified Bessel functions of second kind, Sect.~9.6 in Ref.~\cite{abramowitz1964handbook}, and we defined
\begin{equation}
q=\sqrt{\frac{s}{D}}, \qquad h = \frac{\kappa_{a}}{2\pi a D}.
\end{equation}
Inserting the expression for $\tilde{S}_{\text{irr}}(s\vert r_{0})$ in Eq.~\eqref{eq2Laplace} yields
\begin{equation}
s\tilde{S}_{\text{rev}}(s\vert r_{0}) = 1 - \frac{hqK_{0}(qr_{0})}{(q^{2}+\kappa_{D})K_{1}(qa) + hqK_{0}(qa)},
\end{equation}
where
\begin{equation}
\kappa_{D} = \frac{\kappa_{d}}{D}.
\end{equation}

\begin{figure}
\includegraphics[scale=0.3]{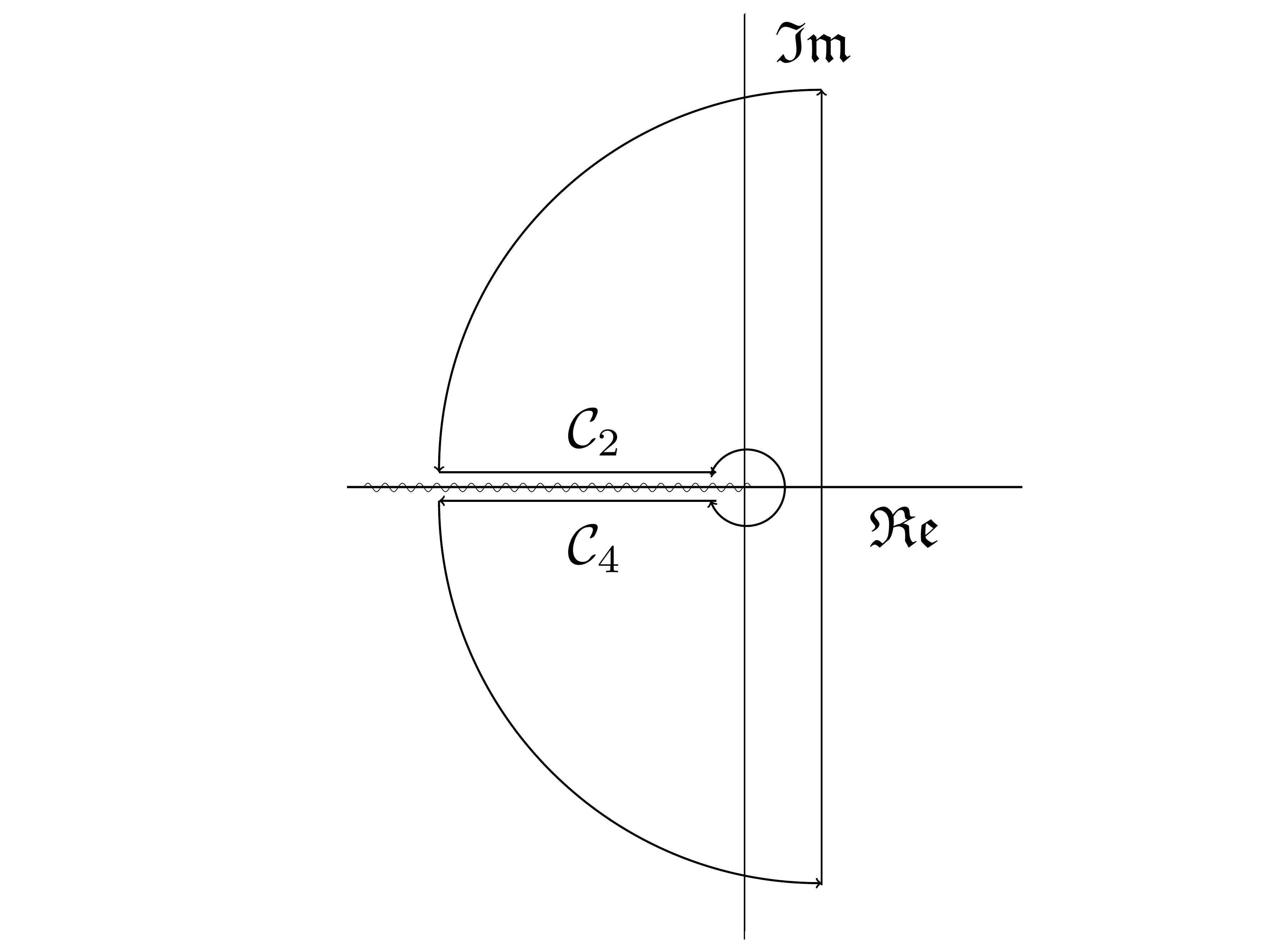}
\caption{Integration contour used  in Eq.~(\ref{cauchy}).}\label{fig:contour}
\end{figure}

The inversion theorem for the Laplace transformation can be applied to find the corresponding expression for $\tilde{S}_{\text{rev}}(s\vert r_{0})$ in the time domain
\begin{equation}\label{inversionFormula}
S_{\text{rev}}(t \vert r_{0}) = \frac{1}{2\pi i} \int^{\gamma+i\infty}_{\gamma-i\infty} e^{st}\,\tilde{S}_{\text{rev}}(s\vert r_{0} )ds.
\end{equation}
To calculate the Bromwich contour integral we first note that $\tilde{S}_{\text{rev}}(s\vert r_{0})$ is multi-valued and has a branch point at $s=0$. Therefore, we use the contour of Fig.~\ref{fig:contour} with a branch cut along the negative real axis, cf. Ch.~12.3, Fig. 40 in Ref.~\cite{carslaw1986conduction}.  Furthermore, we note that the $1/s$ term yields 1 and hence, we obtain 
\begin{eqnarray}\label{cauchy}
2\pi i \big[S_{\text{rev}}(t\vert r_{0}) - 1\big] =  -\int_{\mathcal{C}_{2}} e^{st}\,\tilde{S}_{\text{rev}}(s\vert r_{0} )ds - \int_{\mathcal{C}_{4}} e^{st}\,\tilde{S}_{\text{rev}}(s\vert r_{0} )ds.
\end{eqnarray}
Thus, it remains to calculate the integrals $\int_{\mathcal{C}_{2}}, \int_{\mathcal{C}_{4}}$.
To this end, we choose
$
s = D x^{2} e^{i \pi }
$
and use Append.~3, Eqs.~(25) and (26) in Ref.~\cite{carslaw1986conduction}
\begin{eqnarray}\label{Kwick}
K_{n}(xe^{\pm \pi i/2}) &=& \pm\frac{1}{2}\pi i e^{\mp n\pi i/2} [-J_{n}(x) \pm i Y_{n}(x)],
\end{eqnarray}
where $J_{n}(x), Y_{n}(x)$ denote the Bessel functions of first and second kind, respectively, Sect.~9.1 in Ref.~\cite{abramowitz1964handbook}.
It follows that
\begin{eqnarray}
\int_{\mathcal{C}_{2}}e^{st}\,\tilde{S}_{\text{rev}}(s\vert r_{0})ds  &=&  -2h\int^{\infty}_{0}e^{-Dx^{2}t}[J_{0}(xr_{0})-iY_{0}(xr_{0})]\frac{\alpha(x)+i\beta(x)}{\alpha(x)^{2} + \beta(x)^{2}}   dx. \quad\qquad
\end{eqnarray}
Here we have defined 
\begin{eqnarray}
\alpha(x) &=& ( x^{2} - \kappa_{D})J_{1}(xa) + hxJ_{0}(xa), \label{alphaDef}\\
\beta(x) &=& ( x^{2} - \kappa_{D})Y_{1}(xa) + hxY_{0}(xa).\label{betaDef}
\end{eqnarray}

To evaluate the integral along the contour $\mathcal{C}_{4}$ we choose $s = Dx^{2}e^{-i\pi}$ and after an analogous calculation one finds that
\begin{equation}
\int_{\mathcal{C}_{2}}e^{st}\,\tilde{S}_{\text{rev}}(s\vert r_{0} )ds=-\bigl(\int_{\mathcal{C}_{4}}e^{st}\,\tilde{S}_{\text{rev}}(s\vert r_{0} )ds\bigr)^{\ast},
\end{equation}
where $\ast$ means complex conjugation.
Thus, one finally arrives at the exact expression for the survival probability in the time domain 
\begin{equation}\label{exactS}
S_{\text{rev}}(t\vert r_{0} )=1-a\int^{\infty}_{0}e^{-Dx^{2}t}P(x, a)T(x, r_{0})\,dx,
\end{equation}
where \cite{TPMMS_2012JCP}
\begin{eqnarray}\label{defT}
T(x, r_{0}) &=& \frac{J_{0}(xr_{0})\beta - Y_{0}(xr_{0})\alpha}{[\alpha^{2}+\beta^{2}]^{1/2}}, \\
P(x, r_{0}) &=& -\frac{1}{x}\frac{\partial}{\partial r}T(x, r_{0}) =  \frac{J_{1}(xr_{0})\beta - Y_{1}(xr_{0})\alpha}{[\alpha^{2}+\beta^{2}]^{1/2}}.
\end{eqnarray}

Next, we turn to the case of the initially bound state and use Eqs.~\eqref{SradLaplace},~\eqref{eq1Laplace} to obtain
\begin{equation}\label{exactS*}
\tilde{S}_{\text{rev}}(s\vert \ast) = \frac{\kappa_{D}}{s}\frac{K_{1}(qa)}{(q^{2}+\kappa_{D})K_{1}(qa) + hqK_{0}(qa)}.
\end{equation}
We can use again the inversion theorem Eq.~\eqref{inversionFormula} to calculate the expression for $S_{\text{rev}}(t\vert \ast)$ in the time domain.
The actual calculation is very similar to the one presented for $S_{\text{rev}}(t\vert r_{0})$, therefore we only give the result
\begin{equation}
S_{\text{rev}}(t\vert \ast) = 1-2\pi \frac{\kappa_{d}}{\kappa_{a}}a^{2} \int^{\infty}_{0}e^{-Dx^{2}t}P^{2}(x, a)\frac{dx}{x}.
\end{equation}
A comparison with Ref.~\cite{TPMMS_2012JCP}, where Eqs.~\eqref{exactS} and \eqref{exactS*} were obtained by first calculating the GF, shows that the exact expressions for $S_{\text{rev}}(t\vert \ast), S_{\text{rev}}(t\vert r_{0})$ can be more easily derived via the route of the convolution relations. 
\subsection{Restricted plane} 
Next, we consider the case of the restricted plane. The expression for $S_{\text{irr}}(s\vert r_{0})$ is known, Ch.~13.4., Eqs.~(3) and (4) in Ref.~\cite{carslaw1986conduction}
\begin{eqnarray}\label{SradRestricted}
\tilde{S}_{\text{irr}}(s\vert r_{0}) = \frac{1}{s} + \frac{h}{s}\frac{[K_{1}(qb)I_{0}(qr_{0}) + I_{1}(qb)K_{0}(qr_{0})]}{K_{1}(qb)[qI_{1}(qa) - hI_{0}(qa)] - I_{1}(qb)[qK_{1}(qa) + hK_{0}(qa)]}. \qquad
\end{eqnarray}
Again we employ Eq.~\eqref{eq2Laplace} to arrive at the
survival probability in the Laplace domain
\begin{eqnarray}\label{LaplaceSRestricted}
\tilde{S}_{\text{rev}}(s\vert r_{0}) = \frac{1}{s} + \frac{hq[K_{1}(qb)I_{0}(qr_{0}) + I_{1}(qb)K_{0}(qr_{0})]}{s\Delta(s)},
\end{eqnarray}
where
\begin{equation}
\Delta(s) = \psi(qa)K_{1}(qb) - \phi(qa)I_{1}(qb),
\end{equation}
and
\begin{eqnarray}
\psi(qa) &=& (q^{2}+\kappa_{D})I_{1}(qa) - hqI_{0}(qa)\\
\phi(qa) &=& (q^{2}+\kappa_{D})K_{1}(qa) + hqK_{0}(qa).
\end{eqnarray}
\begin{figure}
\includegraphics[scale=0.3]{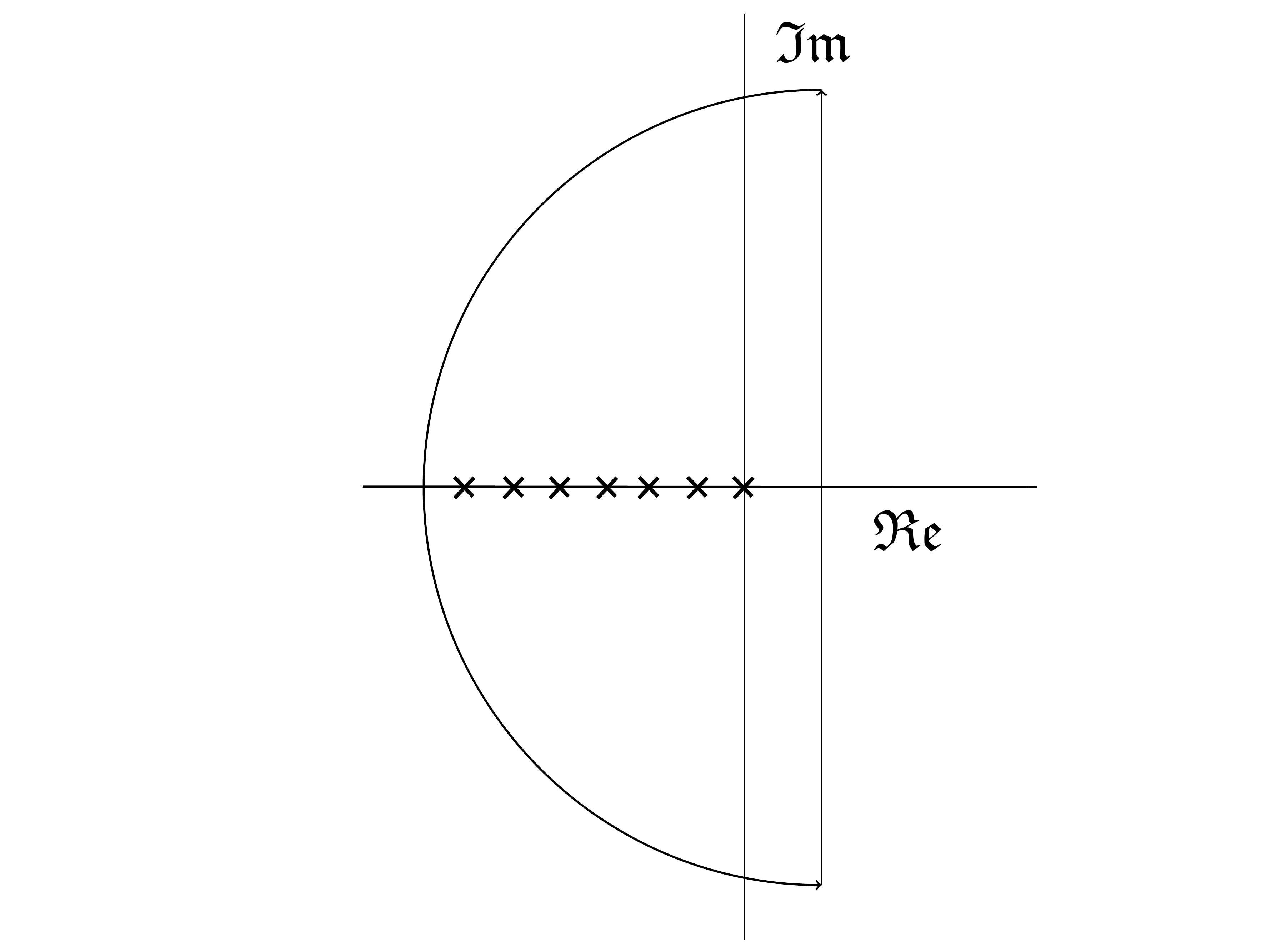}
\caption{Integration contour used in Eq.~(\ref{Sr_rest}).}\label{fig:contour2}
\end{figure}
Next, we again apply the inversion theorem Eq.~\eqref{inversionFormula} to calculate the survival probability in the time domain. In contrast to the case of the infinite plane, the integrand Eq.~\eqref{LaplaceSRestricted}
is now single-valued and we choose the integration contour depicted in Fig. 2. Then, the Bromwich contour integral is determined by   
\begin{equation}
S_{\text{rev}}(t\vert r_{0}) = \frac{1}{2\pi i} \int^{\gamma+i\infty}_{\gamma-i\infty} e^{st}\,\tilde{S}_{\text{rev}}(s\vert r_{0} )ds = \sum_{n}\text{Res}_{s_{n}}\big[e^{st}\tilde{S}_{\text{rev}}(s\vert r_{0})\big],
\end{equation}
where the sum goes over all poles of $e^{st}\tilde{S}_{\text{rev}}(s\vert r_{0})$. The second term in Eq.~\eqref{LaplaceSRestricted} has a simple pole at $s_{0}=0$ and non-vanishing simple poles
at $s_{n}=-D\xi^{2}_{n}$, i.e. one has
\begin{equation}\label{nonZeroPoles}
\Delta(s)\bigg\vert_{s=-D\xi^{2}} = 0,
\end{equation}
where $\pm\xi_{n}\neq 0$ are the roots of 
\begin{equation}\label{realRoots}
\alpha(\xi_{n})Y_{1}(\xi_{n}b) - \beta(\xi_{n})J_{1}(\xi_{n}b) = 0,
\end{equation}  
with $\alpha(x),\beta(x)$ defined by Eqs.~\eqref{alphaDef},~\eqref{betaDef}.
To arrive at Eq.~\eqref{realRoots}, we used Eq.~\eqref{nonZeroPoles} as well as Eq.~\eqref{Kwick} and
\begin{eqnarray}\label{Iwick}
I_{n}(xe^{\pm \pi i/2}) =e^{\pm n\pi i/2} J_{n}(x).
\end{eqnarray}

The residue at the pole $s=0$ can be determined by employing the small argument expansion of the modified Bessel functions, see Sect.~9.6, Eqs. (9.6.7)-(9.6.9) in Ref.~\cite{abramowitz1964handbook}. We obtain
\begin{equation}
\text{Res}_{s=0}\bigg[e^{st}\frac{hq[K_{1}(qb)I_{0}(qr_{0}) + I_{1}(qb)K_{0}(qr_{0})]}{s\Delta(s)}\bigg] = \frac{ha}{\frac{\kappa_{D}}{2}\big(a^{2}-b^{2}\big) -ha}.
\end{equation}
Taking into account the contribution from the $1/s$ term in Eq.~\eqref{LaplaceSRestricted}, we arrive at
\begin{equation}
S(t\vert r_{0}) - \frac{\pi(b^{2}-a^{2})}{\pi(b^{2}-a^{2}) + K_{a}} = \sum_{n\neq 0}\text{Res}_{s_{n}}\bigg[e^{st}\frac{hq[K_{1}(qb)I_{0}(qr_{0}) + I_{1}(qb)K_{0}(qr_{0})]}{s\Delta(s)}\bigg],
\end{equation}
where we introduced the equilibrium constant
\begin{equation}
K_{a}=\frac{\kappa_{a}}{\kappa_{d}}.
\end{equation}
To calculate the remaining residues at the non-vanishing poles $s_{n} = -D\xi^{2}_{n}$, we use
\begin{eqnarray}
&&\text{Res}_{s_{n} = -D\xi^{2}_{n}}\bigg[e^{st}\frac{hq[K_{1}(qb)I_{0}(qr_{0}) + I_{1}(qb)K_{0}(qr_{0})]}{s\Delta(s)}\bigg]\nonumber\\
&&= e^{st}\frac{hq[K_{1}(qb)I_{0}(qr_{0}) + I_{1}(qb)K_{0}(qr_{0})]}{s\frac{d}{ds}\Delta(s)}\bigg\vert_{s=-D\xi^{2}_{n}}.\label{Res}
\end{eqnarray}
The calculation of the denumerator 
\begin{equation}
s \frac{d}{ds}\Delta(s)\bigg\vert_{s=-D\xi^{2}_{n}} = \frac{1}{2}q \frac{d}{dq}\Delta(s)\bigg\vert_{q=i\xi_{n}}
\end{equation}
is greatly facilitated by the identities
\begin{equation}
\frac{\psi(qa)}{I_{1}(qb)}\bigg\vert_{q=i\xi_{n}}= \frac{\phi(qa)}{K_{1}(qb)}\bigg\vert_{q=i\xi_{n}} = -\frac{\alpha(xa)}{J_{1}(xb)}\bigg\vert_{x=\xi_{n}} = -\frac{\beta(xa)}{Y_{1}(xb)}\bigg\vert_{x=\xi_{n}}\equiv\rho,
\end{equation}
which follows from Eqs.~\eqref{nonZeroPoles},~\eqref{Iwick} and \eqref{realRoots}.
Moreover, we make use of the following identities
\begin{eqnarray}
&& xI'_{\nu}(x) + \nu I_{\nu}(x) = xI_{\nu - 1}(x), \nonumber\\
&& xK'_{\nu}(x) + \nu K_{\nu}(x) = -xK_{\nu - 1}(x), \nonumber\\ 
&& I_{\nu}(x)K'_{\nu}(x) - K_{\nu}(x)I'_{\nu}(x) = -\frac{1}{x}, \nonumber\\ 
&& I_{\nu}(x)K_{\nu+1}(x) + K_{\nu}(x)I_{\nu+1}(x) = \frac{1}{x}, 
\end{eqnarray}
and find in this way
\begin{eqnarray}
s \frac{d}{ds}\Delta(s)\vert_{s=-D\xi^{2}_{n}} = \frac{\alpha^{2}(\xi_{n}a) - J^{2}_{1}(\xi_{n}b)\varkappa}{2\alpha(\xi_{n}a)J_{1}(\xi_{n}b)}, 
\end{eqnarray}
where
\begin{equation}
\varkappa = (\xi^{2}_{n} - \kappa_{D})^{2} - 2a^{-1}h\kappa_{D} + h^{2}\xi^{2}_{n} .
\end{equation}
The numerator on the rhs of Eq.~\eqref{Res} can be evaluated by use of Eqs.~\eqref{Kwick},~\eqref{Iwick}.
Everything taken together, we obtain for the survival probability in the restricted plane
\begin{eqnarray}\label{Sr_rest}
S_{\text{rev}}(t\vert r_{0})  = \frac{\pi(b^{2}-a^{2})}{\pi(b^{2}-a^{2}) + K_{a}} - h\pi\sum_{n\neq 0}e^{-D\xi_{n}t}\xi_{n}\frac{C(\xi_{n}r_{0},\xi_{n}b)\alpha(\xi_{n}a)J_{1}(\xi_{n}b)}{\alpha^{2}(\xi_{n}a) - J^{2}_{1}(\xi_{n}b)\varkappa},\qquad
\end{eqnarray}
where
\begin{equation}
C(xr_{0}, xb) = J_{0}(xr_{0})Y_{1}(xb) - Y_{0}(xr_{0})J_{1}(xb).
\end{equation}

Finally, we consider the case of the initially bound state. Eqs.~\eqref{SradRestricted} and \eqref{eq1Laplace} yield
\begin{eqnarray}\label{LaplaceSRestricted*}
\tilde{S}_{\text{rev}}(s\vert \ast) = \frac{\kappa_{D}[K_{1}(qb)I_{1}(qa) - I_{1}(qb)K_{1}(qa)]}{s\Delta(s)}.
\end{eqnarray}
The corresponding expression in the time domain can be found along similar lines as presented for $S_{\text{rev}}(t\vert r_{0})$, therefore we only provide the result
\begin{eqnarray}\label{Sast_rest}
S_{\text{rev}}(t\vert \ast)  = \frac{\pi(b^{2}-a^{2})}{\pi(b^{2}-a^{2}) + K_{a}} - \kappa_{D}\pi\sum_{n\neq 0}e^{-D\xi_{n}t}\frac{G(\xi_{n}a,\xi_{n}b)\alpha(\xi_{n}a)J_{1}(\xi_{n}b)}{\alpha^{2}(\xi_{n}a) - J^{2}_{1}(\xi_{n}b)\varkappa},\qquad
\end{eqnarray}
where
\begin{equation}
G(xr_{0}, xb) = J_{1}(xr_{0})Y_{1}(xb) - Y_{1}(xr_{0})J_{1}(xb) = -\frac{1}{x}\frac{\partial}{\partial r_{0}}C(xr_{0}, xb).
\end{equation}
\subsection{Spherical surface}
\begin{figure}
\includegraphics[scale=0.3]{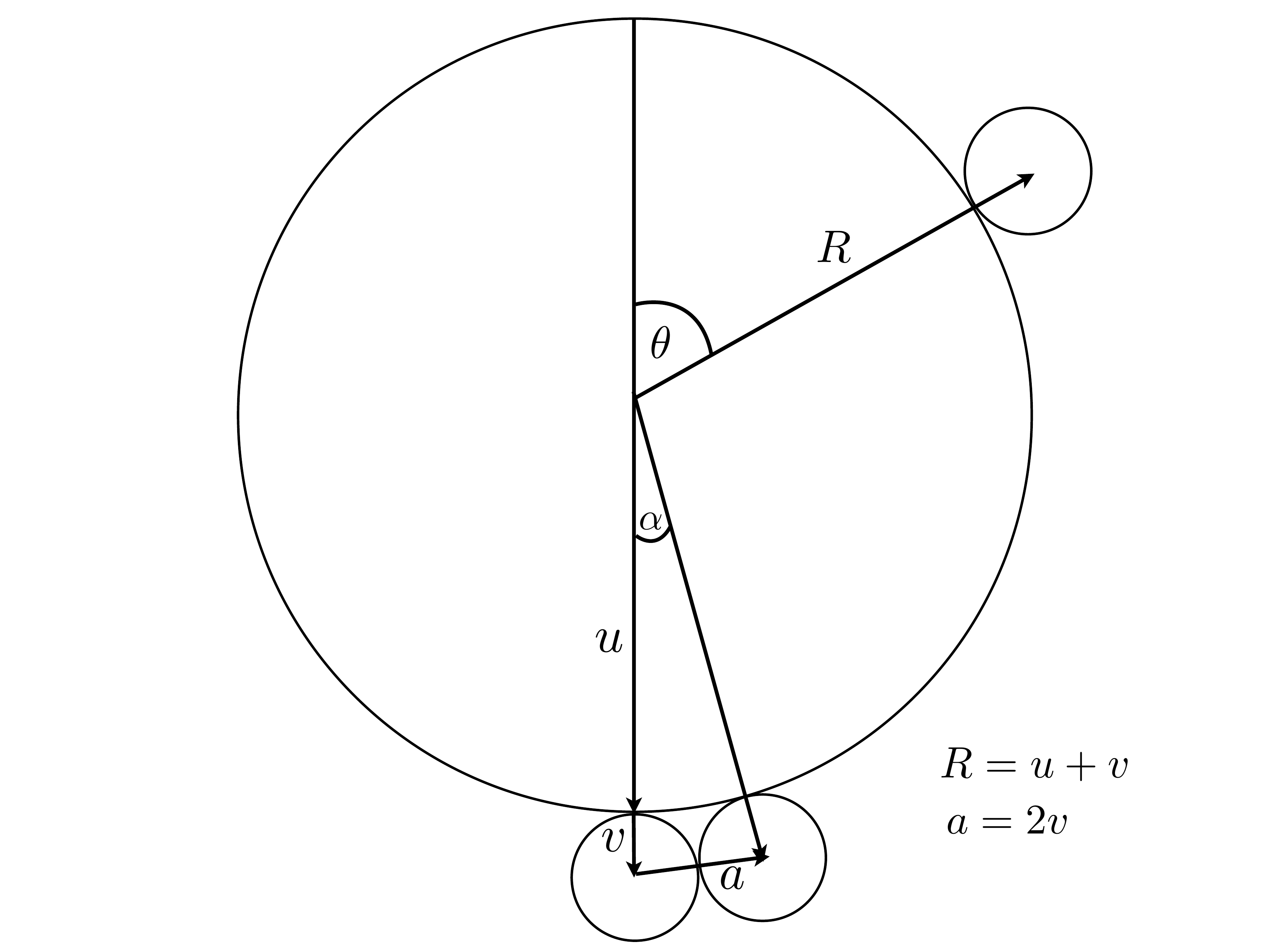}
\caption{Geometry of an isolated pair of molecules on a spherical surface.}\label{fig:Circle}
\end{figure}

Finally, we consider an isolated pair of sphere-like particles with encounter radius $a$ diffusing on the surface of a sphere of radius $u$, cf. Ref.~\cite{Sano_Tachiya:1981}. Without loss of generality, one particle can be thought of as fixed 
at the south pole $\theta = \pi$, while the other particle diffuses
around characterized by the diffusion constant $D=D_{A}+D_{B}$. The position of the diffusing particle is given by the angle $\theta$, or equivalently by $z=\cos \theta$. The particles are at contact if $z = z_{a} \equiv -\cos\alpha$, see Fig.~\ref{fig:Circle}. 
By making use of the S-T equation, the survival probability in the irreversible case has been calculated in Ref.~\cite{Sano_Tachiya:1981}
\begin{equation}\label{SradSphere}
S_{\text{irr}}(t\vert z) = \sum^{\infty}_{n=1}c_{i}e^{-t/\tau_{i}}P_{\nu_{i}}(z), 
\end{equation} 
where the functions $P_{\nu_{i}}$ denote those fundamental solutions of the Legendre differential equation that are finite at $z=1$.
The numbers $\nu_{i}$ are the roots of the radiation boundary condition of the S-T equation. Furthermore, we have
\begin{equation}
\tau_{i} = \frac{\tau}{\nu_{i}(\nu_{i}+1)},\quad\tau = \frac{R^{2}}{D},
\end{equation}
and
\begin{equation}
c_{i} = \frac{\int^{1}_{-\cos\alpha}P_{\nu_{i}}(z)dz}{\int^{1}_{-\cos\alpha} P^{2}_{\nu_{i}}(z)dz}.
\end{equation}
Using the Laplace transform of Eq.~\eqref{SradSphere}
\begin{equation}\label{SradSphereLaplace}
\tilde{S}_{\text{irr}}(s\vert z) = \sum^{\infty}_{n=1}\frac{c_{i}P_{\nu_{i}}(z)}{s+\tau^{-1}_{i}}, 
\end{equation}
and using Eqs.~\eqref{eq1Laplace} and~\eqref{eq2Laplace}, we can immediately obtain exact expressions for $\tilde{S}_{\text{rev}}(s\vert z), \tilde{S}_{\text{rev}}(s\vert \ast)$ 
in the Laplace domain. 

However, to derive expressions in the time domain we will take another route.
As described in more detail in Ref.~\cite{Sano_Tachiya:1981}, the survival probability in a restricted space is typically given by a sum of exponentials that can be approximated by a single
exponential, with the exception of very short times
\begin{equation}\label{Sapprox}
S_{\text{irr,app}}(t\vert z) = \exp\bigg[-\frac{t}{\tau(z)}\bigg],
\end{equation}
where the decay time $\tau(z)$ may be identified with the mean reaction time, which is given by \cite{Sano_Tachiya:1981}
\begin{equation}\label{defMRT}
\tau(z) = \int^{\infty}_{0}t\frac{\partial}{\partial t}\big[1-S_{\text{irr}}(t\vert z)\big]dt = \int^{\infty}_{0}S_{\text{irr}}(t\vert z)dt.
\end{equation}
For the case of a pair on a spherical surface, the mean reaction time has been calculated in Ref.~\cite{Sano_Tachiya:1981}.
Now, the Laplace transform of the approximative expression Eq.~\eqref{Sapprox} is very simple
\begin{equation}
\tilde{S}_{\text{irr, app}}(s\vert z) = \frac{1}{s+\tau(z)^{-1}},
\end{equation}
and via Eqs.~\eqref{eq1Laplace},~\eqref{eq2Laplace} we immediately obtain
\begin{eqnarray}
\tilde{S}_{\text{rev,app}}(s\vert \ast) & = & \frac{\kappa_{d}}{s}\frac{1}{s+\tau^{-1}(z_{a}) + \kappa_{d}}, \\
\tilde{S}_{\text{rev,app}}(s\vert z) & = & \frac{1}{s + \tau^{-1}(z)}\bigg[1 - \frac{\kappa_{d}}{s+\tau^{-1}(z_{a}) + \kappa_{d}}\bigg]  \nonumber\\
&&+ \tilde{S}_{\text{rev,app}}(s\vert \ast),
\end{eqnarray}
where $z_{a} = -\cos\alpha$.
Then, the corresponding expressions in the time domain become
\begin{eqnarray}
S_{\text{rev,app}}(t\vert \ast) & = & \frac{\kappa_{d}}{\kappa_{d} + \tau^{-1}(z_{a})}\bigg[1 - e^{-t[\tau^{-1}(z_{a})+\kappa_{d}]}\bigg], \label{SrevSphereStar}\\
S_{\text{rev,app}}(t\vert z) & = & \frac{\kappa_{d}}{\kappa_{d} + \tau^{-1}(z_{a})}  + e^{-t/\tau(z)}\frac{\tau^{-1}(z) - \tau^{-1}(z_{a})}{\tau^{-1}(z) - \tau^{-1}(z_{a}) - \kappa_{d}} \nonumber\\
&&- e^{-t[\kappa_{d}+\tau(z_{a})^{-1}]}\bigg[\frac{\kappa_{d}}{\kappa_{d} + \tau^{-1}(z_{a})} + \frac{\kappa_{d}}{\tau^{-1}(z) - \tau^{-1}(z_{a}) - \kappa_{d}}\bigg].  \label{SrevSphereR}\qquad\qquad
\end{eqnarray}
Note that
\begin{eqnarray}
S_{\text{rev, app}}(t = 0\vert \ast) &=& 0, \\
S_{\text{rev,app}}(t = 0\vert z) &=& 1,
\end{eqnarray}
as it should be. Furthermore, for the steady-state we obtain
\begin{equation}
\lim_{t\rightarrow\infty}S_{\text{rev,app}}(t\vert \ast) = \lim_{t\rightarrow\infty}S_{\text{rev,app}}(t \vert z) = \frac{\kappa_{d}}{\kappa_{d} + \tau^{-1}(z_{a})},
\end{equation}
and find that the steady-state is independent of the initial state, as expected.

It is now instructive to calculate $S_{\text{rev,app}}(t\vert r_{0}), S_{\text{rev,app}}(t\vert \ast)$ for the restricted plane and to compare the results with the case of a sphere. 
To this end, we first note that the approximate expressions Eqs.~\eqref{SrevSphereStar},~\eqref{SrevSphereR} for the survival probabilities are still valid also in the case of the restricted plane, due to the general nature of the relations Eqs.~\eqref{eq1},~\eqref{eq1unbound} and the approximation Eq.~\eqref{Sapprox}. It only remains to calculate the mean passage time $\tau(r_{0})$ for the case of the restricted plane. By means of
Eq.~\eqref{defMRT} it follows that 
\begin{equation}
 \tau(r_{0}) = \lim_{s\rightarrow 0} \tilde{S}_{\text{irr}}(s\vert r_{0}).
\end{equation}      
Using Eq.~\eqref{SradRestricted} and the series expansion of $I_{\nu}$ and $K_{\nu}$, Sect.~9.6, Eqs.~(9.6.10) - (9.6.11) in Ref.~\cite{abramowitz1964handbook}, we obtain
\begin{equation}\label{MRT}
 \tau^{\text{P}}(r_{0}) = \frac{b^{2}}{2D}\ln(r_{0}/a) - \frac{1}{4D}(r_{0}^{2}-a^{2}) +\frac{1}{2haD}(b^{2}-a^{2}).
\end{equation}
There is an alternative way of deriving Eq.~\eqref{MRT}. The mean reaction time $\tau(r)$ is related to $S_{\text{irr}}(t \vert r_{0})$ through Eq.~\eqref{defMRT}. On the other hand, $S_{\text{irr}}(t \vert r_{0})$ satisfies the Sano-Tachiya equation. Therefore, the mean reaction time is calculated from the differential equation and the associated boundary conditions obtained by integrating the Sano-Tachiya equation over $t$ from $0$ to infinity \cite{Sano_Tachiya:1981, SzaboSchulten:1980}. 

Henceforth, we will use $\tau^{\text{P}}(r_{0}), \tau^{\text{S}}(z)$ to denote the mean reaction time for the case of the restricted plane and of the sphere, respectively. 
The mean reaction time of the sphere has been given in Ref.~\cite{Sano_Tachiya:1981}. In the notation of the present article, one has at contact, $z  =-\cos\alpha = -1+\frac{a^{2}}{2 R^{2}}$
\begin{equation}\label{MRTC}
\tau^{\text{S}}(z = -\cos\alpha) = 4\pi\frac{R^{2}}{\kappa_{a}}\bigg [1 - \frac{a^{2}}{4R^{2}}\bigg]^{1/2}.
\end{equation} 
Now, to facilitate a comparison between the two cases, we follow Ref.~\cite{Sano_Tachiya:1981} and consider a pair of molecules in a circle of radius $b=2R$ such that the area available for diffusion
is equal to that on the surface of a sphere with radius $R$. Then, by virtue of Eq.~\eqref{MRT} the mean reaction time becomes at contact $r_{0}=a$
\begin{equation}\label{MRTC2}
\tau^{\text{P}}(r_{0}=a) = 4\pi\frac{R^{2}}{\kappa_{a}}\bigg [1 - \frac{a^{2}}{4R^{2}}\bigg].
\end{equation}  
Clearly, one has
\begin{equation}
\tau^{\text{S}}(z = -\cos\alpha) > \tau^{\text{P}}(r_{0}=a).
\end{equation}
Using this relation and Eqs.~\eqref{SrevSphereStar}, ~\eqref{SrevSphereR}, we can compare the survival probabilities in the case of the restricted plane and the sphere.
In particular, we find for the ultimate fate of the molecule pair
\begin{equation}\label{ulti}
\lim_{t\rightarrow\infty}S^{\text{S}}_{\text{rev,app}}(t\vert \ast) > \lim_{t\rightarrow\infty}S^{\text{P}}_{\text{rev,app}}(t\vert \ast). 
\end{equation} 
\section{Concluding remarks}
In this article, we investigated the reversible diffusion-influenced reaction of an isolated pair in two dimensions for the case of an infinite and restricted plane and the surface of a sphere.
As the central theoretical tool we employed the convolution relations Eqs.~\eqref{eq1},~\eqref{eq1unbound} that permit to express the survival probability of the reversible reaction directly in terms of the survival probability of the irreversible reaction, which is already known for many cases. Compared to approaches that involve the explicit derivation of the GF of the underlying Smoluchowski equation, the discussed method considerably reduces the complexity of the necessary calculations as demonstrated by the derivation of the exact expressions Eqs.~\eqref{exactS},~\eqref{exactS*},~\eqref{Sr_rest} and \eqref{Sast_rest} for the case of the infinite and restricted plane, respectively. 
Moreover, we combined the approach based on the convolution relations with the mean reaction time approximation method that gives the irreversible survival probability in restricted spaces as a single exponential Eq.~\eqref{Sapprox}. Thus, we derived approximate expressions in the time domain for the reversible survival probability for the case of the surface of a sphere, Eqs.~\eqref{SrevSphereStar},~\eqref{SrevSphereR}. We derived analogue approximate expressions for the survival probability in the case of the restricted plane. In this way, we could readily compare the influence of the underlying diffusion spaces on the behavior of the reversible diffusion-influenced system. In particular, we found that the ultimate separation probability of an isolated pair is larger in the case of a sphere than in the case of a restricted plane Eq.~\eqref{ulti}.    
\section*{Appendix}\label{appendix:Dyson}
In this appendix, we present an alternative derivation of the convolution equations Eqs.~\eqref{eq1},~\eqref{eq2} and ~\eqref{eq1unbound},~\eqref{eq2unbound}.
Without loss of generality, we consider the case of the infinitely extended plane.

We start by considering the reversible GF $p_{\text{rev}}(r,t\vert r_{0})$ that satisfies
\begin{equation}\label{app:diffusionEq}
\frac{\partial p_{\text{rev}}(r, t\vert r_{0})}{\partial t} = \mathcal{L}_{r}p_{\text{rev}}(r, t\vert r_{0}),
\end{equation}
where the differential operator $\mathcal{L}_{r}$ is defined by
\begin{equation}
\mathcal{L}_{r} = D\frac{1}{r}\frac{\partial}{\partial r} r \frac{\partial}{\partial r}. 
\end{equation}
Obviously, Eq.~\eqref{app:diffusionEq} is equivalent to Eq.~\eqref{diffusionEq}. In addition, we consider the irreversible GF $p_{\text{irr}}(r,t\vert r_{0})$ that also satisfies the Smoluchowski Eq.~\eqref{app:diffusionEq} equation and that is subject to the radiation BC Eq.~\eqref{RadBC}, which describes an irreversible association, instead of the back-reaction BC Eq.~\eqref{BRBC}. 
Also, both GF satisfy the same initial condition Eq.~\eqref{initial_bc}.

For the irreversible GF, we also consider the adjoint form of the Smoluchowski equation and of the radiation BC, i.e.
\begin{equation}\label{app:adjointDiffEq}
-\frac{\partial p_{\text{irr}}(r', t'\vert r, t)}{\partial t} = \mathcal{L}_{r}p_{\text{irr}}(r', t'\vert r, t),
\end{equation}
and
\begin{equation}\label{abc}
2\pi a D \frac{\partial p_{\text{irr}}(r',t'\vert r, t)}{\partial r}\bigg\vert_{r = a} =\kappa_{a}p_{\text{irr}}(r',t'\vert a, t). 
\end{equation}
Now, we multiply Eq.~\eqref{app:diffusionEq} with $p_{\text{irr}}(r',t'\vert r, t)$ and Eq.~\eqref{app:adjointDiffEq} with $p_{\text{rev}}(r, t\vert r_{0})$ and subtract the resulting equations from each other.
The result is
\begin{eqnarray}\label{mergedEq}
\frac{\partial}{\partial t}\bigg[p_{\text{irr}}(r',t'\vert r, t)p_{\text{rev}}(r, t \vert r_{0})\bigg] &=& D\frac{1}{r}\frac{\partial}{\partial r}\bigg[p_{\text{irr}}(r',t'\vert r, t)r\frac{\partial}{\partial r}p_{\text{rev}}(r, t \vert r_{0})  \nonumber\\
&- &p_{\text{rev}}(r, t \vert r_{0}) r\frac{\partial}{\partial r}p_{\text{irr}}(r',t'\vert r, t) \bigg].
\end{eqnarray}
Next, we integrate both sides of Eq.~\eqref{mergedEq} over time $\int^{t'}_{0}dt$ and over space $2\pi\int^{\infty}_{a}dr\, r$.

First, we focus on the lhs of Eq.~\eqref{mergedEq}. After the integration over time it becomes
\begin{equation}
\frac{1}{2\pi r}\bigg[\delta(r' - r)p_{\text{rev}}(r, t' \vert r_{0}) - \delta(r - r_{0})p_{\text{irr}}(r', t' \vert r)\bigg],
\end{equation}
where we have used the initial condition Eq.~\eqref{initial_bc}. The integration over space is now trivial and we finally arrive at
\begin{equation}
2\pi\int^{\infty}_{a}dr\, r\int^{t'}_{0}dt \frac{\partial}{\partial t}\bigg[p_{\text{irr}}(r',t'\vert r)p_{\text{rev}}(r, t \vert r_{0})\bigg] = p_{\text{rev}}(r', t' \vert r_{0}) - p_{\text{irr}}(r', t' \vert r_{0}).
\end{equation}
Now, we turn to the rhs of Eq.~\eqref{mergedEq}. We integrate first over space and obtain the expression
\begin{equation}
-2\pi a D\bigg[p_{\text{irr}}(r',t'\vert a, t)\frac{\partial}{\partial r}p_{\text{rev}}(r, t \vert r_{0})\bigg\vert_{r = a} - p_{\text{rev}}(a, t \vert r_{0}) \frac{\partial}{\partial r}p_{\text{irr}}(r',t'\vert r, t)\bigg\vert_{r = a}\bigg],
\end{equation}
because the GF vanish for $r\rightarrow\infty$. 
Using the boundary conditions Eqs.~\eqref{BRBC} and~\eqref{abc} we arrive at
\begin{equation}
2\pi\int^{\infty}_{a}dr r\int^{t'}_{0}dt \,[\text{rhs of Eq.~\eqref{mergedEq}}] = \kappa_{d}\int^{t'}_{0}dt[1-S_{\text{rev}}(t\vert r_{0})]p_{\text{irr}}(r',t'\vert a, t).
\end{equation}
Everything taken together, we finally obtain
\begin{equation}\label{preDyson}
p_{\text{rev}}(r', t' \vert r_{0}) - p_{\text{irr}}(r', t' \vert r_{0}) = \kappa_{d}\int^{t'}_{0}dt[1-S_{\text{rev}}(t\vert r_{0})]p_{\text{irr}}(r',t'\vert a, t).
\end{equation}
We note that 
\begin{equation}
p_{\text{irr}}(r',t'\vert a, t) = p_{\text{irr}}(r',t'-t\vert a).
\end{equation}
Using this identity and switching the notation $t\leftrightarrow t', r'\rightarrow r$, Eq.~\eqref{preDyson} takes the form
\begin{equation}\label{Dyson}
p_{\text{rev}}(r, t \vert r_{0}) = p_{\text{irr}}(r, t \vert r_{0}) + \kappa_{d}\int^{t}_{0}dt'[1-S_{\text{rev}}(t'\vert r_{0})]p_{\text{irr}}(r,t-t'\vert a).
\end{equation}
We will discuss elsewhere that Eq.~\eqref{Dyson} is the Dyson equation connecting the reversible GF $p_{\text{rev}}(r, t \vert r_{0})$ subject to a back-reaction BC Eq.~\eqref{BRBC} with the irreversible GF $p_{\text{irr}}(r, t \vert r_{0})$ subject to a radiation BC Eq.~\eqref{RadBC}. Dyson equations relating the non-reactive GF subject to a reflective BC with $p_{\text{irr}}(r, t \vert r_{0})$ and $p_{\text{rev}}(r, t \vert r_{0})$ have been considered in Refs.~\cite{SzaboLammWeiss:1984} and \cite{GopichAgmon:1999, TPMMS_2012JCP}. An useful feature of the Dyson equation Eq.~\eqref{Dyson} is that it gives rise to the convolution relations, as we shall see now.  

In fact, by integrating over space $2\pi\int^{\infty}_{a}dr r$ and using the definition Eq.~\eqref{defS}, the convolution relation Eq.~\eqref{eq1unbound} immediately follows from the Dyson equation Eq.~\eqref{Dyson}  
\begin{equation}\label{eq1v2}
S_{\text{rev}}(t\vert r_{0}) = S_{\text{irr}}(t\vert r_{0}) + \kappa_{d}\int^{t}_{0}[1-S_{\text{rev}}(t'\vert r_{0})]S_{\text{irr}}(t-t'\vert a) dt'.
\end{equation}

Next, we differentiate both sides of Eq.~\eqref{eq1v2} wrt time and obtain
\begin{eqnarray}
&&\frac{\partial S_{\text{rev}}(t\vert r_{0})}{\partial t} = \frac{\partial S_{\text{irr}}(t\vert r_{0})}{\partial t} + \kappa_{d}[1-S_{\text{rev}}( t\vert r_{0})]S_{\text{irr}}(0\vert a) \nonumber\\
&&+ \kappa_{d}\int^{t}_{0}[1-S_{\text{rev}}(t'\vert r_{0})]\frac{\partial S_{\text{irr}}(t-t'\vert a)}{\partial t} dt' \nonumber\\
&&= \kappa_{d}P_{\text{rev}}( t\vert r_{0}) + \kappa_{d}\int^{t}_{0}P_{\text{rev}}(t'\vert r_{0})\frac{\partial S_{\text{irr}}(t-t'\vert a)}{\partial t} dt' + \frac{\partial S_{\text{irr}}(t\vert r_{0})}{\partial t},\nonumber
\end{eqnarray}
where we used $S_{\text{irr}}(0\vert a) = 1$.
Obviously,
\begin{equation}
\frac{\partial S_{\text{rev}}(t\vert r_{0})}{\partial t} = -\frac{\partial P_{\text{rev}}(t\vert r_{0})}{\partial t},
\end{equation}
and hence Eq.~\eqref{eq2unbound} follows.

Finally, we would like to point out that Eqs.~\eqref{eq1} and \eqref{eq2} can be obtained along the same lines. The major difference is that the reversible GF $p_{\text{rev}}(r, t \vert \ast)$ satisfies the initial condition Eq.~\eqref{initialBCBound} instead of Eq.~\eqref{initial_bc}. This difference leads to the following Dyson equation
\begin{equation}\label{Dyson}
p_{\text{rev}}(r, t \vert \ast) = \kappa_{d}\int^{t}_{0}dt'[1-S_{\text{rev}}(t'\vert \ast)]p_{\text{irr}}(r,t-t'\vert a),
\end{equation}
and hence Eqs.~\eqref{eq1} and \eqref{eq2} follow.
\subsection*{Acknowledgments}
This research was supported in part by the Intramural Research Program of the NIH, National Institute of Allergy and Infectious Diseases. \newline
T. P. would like to thank Martin Meier-Schellersheim for helpful comments. 
\bibliographystyle{ieeetr}

\begin{thebibliography}{10}

\bibitem{Rice:1985}
S.~A. Rice, {\em Diffusion Limited Reactions}.
\newblock New York: Elsevier, 1985.

\bibitem{Agmon:1984}
N.~Agmon {\em J. Chem. Phys.}, vol.~81, p.~2811, 1984.

\bibitem{kimShin:1999}
H.~Kim and K.~Shin {\em Phys. Rev. Lett.}, vol.~82, p.~1578, 1999.

\bibitem{TPMMS_2012JCP}
T.~Pr\"ustel and M.~Meier-Schellersheim {\em J. Chem. Phys.}, vol.~137,
  p.~054104, 2012.

\bibitem{Tachiya:1980}
M.~Tachiya, ``Theory of diffusion-controlled dissociation and its applications
  to charge separation.'' Extended abstract of Annual Meeting on
  Photochemistry, p. 256-257, Tsu, Japan, 1980.

\bibitem{Agmon:1990p10}
N.~Agmon and A.~Szabo {\em J. Chem. Phys.}, vol.~92, p.~5270, 1990.

\bibitem{Sano_Tachiya:1979}
H.~Sano and M.~Tachiya {\em J. Chem. Phys.}, vol.~71, p.~1276, 1979.

\bibitem{Sano_Tachiya:1981}
H.~Sano and M.~Tachiya {\em J. Chem. Phys.}, vol.~75, p.~2870, 1981.

\bibitem{Toussaint_Wilczek_1983}
D.~Toussaint and F.~Wilczek {\em J. Chem. Phys.}, vol.~78, p.~2642, 1983.

\bibitem{Emeis_Fehder_1970}
C.~Emeis and P.~Fehder {\em J. Am. Chem. Soc.}, vol.~92, p.~2246, 1970.

\bibitem{bethani:2010}
I.~Bethani, S.~Skanland, I.~Dikic, and A.~Acker-Palmer {\em EMBO J.}, vol.~29,
  p.~2677, 2010.

\bibitem{TPMMS_2013JCP}
T.~Pr\"ustel and M.~Meier-Schellersheim {\em J. Chem. Phys.}, vol.~138,
  p.~104112, 2013.

\bibitem{smoluchowski:1917}
M.~von Smoluchowski {\em Z. Phys. Chem.}, vol.~92, p.~129, 1917.

\bibitem{collins1949diffusion}
F.~Collins and G.~Kimball {\em J. Colloid Sci.}, vol.~4, p.~425, 1949.

\bibitem{carslaw1986conduction}
H.~Carslaw and J.~Jaeger, {\em Conduction of Heat in Solids}.
\newblock New York: Clarendon Press, 1986.

\bibitem{abramowitz1964handbook}
M.~Abramowitz and I.~Stegun, {\em Handbook of Mathematical Functions with
  Formulas, Graphs, and Mathematical Tables}.
\newblock New York: Dover, 1965.

\bibitem{SzaboSchulten:1980}
A.~Szabo, K.~Schulten, and Z.~Schulten {\em J. Chem. Phys.}, vol.~72, p.~4350,
  1980.

\bibitem{SzaboLammWeiss:1984}
A.~Szabo, G.~Lamm, and G.~Weiss {\em J. Stat. Phys.}, vol.~34, p.~225, 1984.

\bibitem{GopichAgmon:1999}
I.~Gopich, K.~Solntsev, and N.~Agmon {\em J. Chem. Phys.}, vol.~110, p.~2164,
  1999.

\end{thebibliography}

\end{document}